\documentclass[12pt]{article}
\usepackage{epsfig}

\newcommand{\pard}[2]{\frac{\partial #1}{\partial #2}}

\def \exp {\mathop{\rm exp}\nolimits}
\def \pq  {\mathop{\rm pq}\nolimits}
\def \ps  {\mathop{\rm ps}\nolimits}

\def \ds  {\mathop{\rm ds}\nolimits}
\def \ns  {\mathop{\rm ns}\nolimits}
\def \cs  {\mathop{\rm cs}\nolimits}
\def \sn  {\mathop{\rm sn}\nolimits}
\def \cn  {\mathop{\rm cn}\nolimits}
\def \dn  {\mathop{\rm dn}\nolimits}

\def \ha  {\mathop{\rm h}\nolimits}

\def \sech{\mathop{\rm sech}\nolimits}

\def \Sign {\sigma}
\def \ax {a_x} \def \at {a_t} \def \kx {k_x} \def \kt {k_t}
\def \ax {r}   \def \at {s}   \def \kx {k}   \def \kt {k_1}
\def \taux {\tau}   \def \taut {\tau_1}

\begin{document}

\title{Analytic doubly periodic wave patterns for the integrable discrete
nonlinear Schr\"odinger (Ablowitz-Ladik) model\footnote
{Corresponding author K.W.~Chow, phone +852--28592641, fax +852--28585415.
Preprint S2005/013.  nlin.PS/0509005} 
}

\author{K.W.~Chow\dag,\ Robert Conte\ddag\ and Neil Xu$^\circ$
{}\\
\\ \dag Department of Mechanical Engineering, University of Hong Kong
\\ Pokfulam, Hong Kong
\\ E-mail: kwchow@hkusua.hku.hk
{}\\
\\ \ddag Service de physique de l'\'etat condens\'e (URA 2464), CEA--Saclay
\\ F--91191 Gif-sur-Yvette Cedex, France
\\ E-mail:  Conte@drecam.saclay.cea.fr
{}\\
\\ $^\circ$ Department of Mathematics, California Institute of Technology
\\ Pasadena, CA 91125, USA
\\ E-mail:  naijie@caltech.edu
}

\maketitle

\hfill 

{\vglue -10.0 truemm}
{\vskip -10.0 truemm}

\begin{abstract}
We derive two new solutions in terms of elliptic functions,
one for the dark and one for the bright soliton regime,
for the semi-discrete
cubic nonlinear Schr\"odinger equation of Ablowitz and Ladik.
When considered in the complex plane, these two solutions are identical.
In the continuum limit,
they reduce to known elliptic function solutions.
In the long wave limit,
the dark one reduces to the collision of two discrete dark solitons,
and the bright one to a discrete breather.
\end{abstract}


\noindent \textit{PACS 2001.}
  02.30.Jr,
  02.30.Ik,
  42.65.Wi.

\baselineskip=12truept 

\tableofcontents

\section{Introduction}

\def \ax {a_x} \def \at {a_t} \def \kx {k_x} \def \kt {k_t}
\def \ax {r}   \def \at {s}   \def \kx {k}   \def \kt {k_1}
\indent
\phantom{xx}
\indent
Discrete versions of the nonlinear Schr\"odinger (NLS) systems have received
tremendous attention recently,
as this family of evolution equations has been demonstrated to be widely
applicable to many physical disciplines.
Although straightforward discretizations of NLS using second order
central difference are usually nonintegrable,
these models have served as an illustration of the relationship between
disorder and nonlinearity.
From the perspective of lattice dynamics,
such studies are relevant in fields like biology, condensed matter physics,
fiber optics,
and material science \cite{BKRK,KKK,KKT}.
As examples for discussion, the Raman scattering spectra of an electronic material,
the irreversible delocalizing transition of Bose-Einstein condensates trapped
in two dimensional optical lattices and optical fibers,
instabilities in coupled arrays of waveguides are all relevant applications
of the discrete NLS model \cite{BKRK}.
In other applications,
localized impurities in physical systems can introduce wave scattering phenomena,
and excite peculiar modes at the impurity sites.
Lattice models with repulsive nonlinear defects 
incorporating quintic nonlinearities
have been studied \cite{KKK}.

A huge variety of approximation methods,
e.g., finite difference schemes and variational techniques,
has been employed for these discrete evolution equations of NLS type \cite{BKRK,KKT}.
The Ablowitz-Ladik model \cite{AblowitzLadik}
is an important exception where substantial
analytical progress can be made,
as explicit expressions for solitons and breathers can be found
\cite{CRG1995,CRG1996}.

Discrete breathers (DBs),
used loosely here to denote time periodic oscillations in a localized domain,
have been studied intensively.
DBs of the discrete NLS can interact with internal modes
or standing wave phonons and exhibit a rich set of dynamics \cite{JA}.
DBs are also important in the consideration of self focusing and collapse phenomena.

The focus of the present work is the special,
integrable discretization of the NLS represented by
the Ablowitz-Ladik (AL) model \cite{AOT},
\begin{eqnarray}
& &
i \pard{A_n}{t} + \frac{A_{n+1}+A_{n-1}-2 A_n}{h^2}
  + \Sign A_n A_n^*    (A_{n+1}+A_{n-1})=0,\
  \Sign^2=1,
\label{eqNLSDiscreteAL}
\end{eqnarray}
with the notation $A_n\equiv A(x,t), x= n h$.
Analytical advances for the AL and related models
\cite{AOT,GuptaLee,MY,MOJ,KKT}
will be desirable, both as useful, basic knowledge on discrete evolution
equations, as well as providing insight for the critically important
nonintegrable cases.
As usual,
the cases $\Sign>0$ and $\Sign<0$ in (\ref{eqNLSDiscreteAL}) correspond to,
respectively,
the focusing (bright soliton) regime,
and
the defocusing (dark soliton) regime.
The main contribution here is to show that several families of exact solutions
for the continuous NLS have their counterparts in the discrete versions
and to provide these analytic expressions.
Solitary waves,
i.e.~reductions $(x,t) \to x-ct$,
have been studied earlier in the literature,
and hence attention is devoted here to other solutions.
To be precise, we start with the doubly periodic (periodic in both $x$ and $t$)
solutions for the continuous NLS which can be expressed as rational functions of
elliptic or theta functions.
For simplicity, we shall call them \textit{bi-elliptic} solutions.
We choose, as an illustrative example, the solution
\cite{MLB1993,AA1993,Chow2002}
\begin{eqnarray}
& & \left\lbrace
\begin{array}{ll}
\displaystyle{
A=\frac{\ax \kt}{\sqrt{1+\kt}}
\left\lbrack
\frac{\cn(\at t,\kt) + i \sqrt{1+\kt} \dn(\ax x,\kx) \sn(\at t,\kt)}
     {\dn(\at t,\kt) +   \sqrt{1+\kt} \dn(\ax x,\kx)}
\right\rbrack
e^{-i \Omega t},
}
\\
\displaystyle{
\Omega=\frac{2 \ax^2}{1+\kt},
\kx^2 =\frac{2 \kt}  {1+\kt},\
\at   =\frac{2 \ax^2}{1+\kt},\
}
\end{array}
\right.
\label{eqbielliptic_continuous_sncndn_dark}
\end{eqnarray}
in which the four parameters
$\kx,\kt$ (the distinct moduli of the Jacobi elliptic functions)
and $\ax,\at$ (the wave numbers or periods in $x$ and $t$)
are constrained by the two indicated relations.
In theta functions notations, an equivalent representation is
\begin{eqnarray}
& & {\hskip -10.0truemm} \left\lbrace
\begin{array}{ll}
\displaystyle{
A=\frac
{\alpha \theta_3(0,\taux)\theta_4(0,\taux)\theta_2(0,\taut)}
{\theta_4(0,\taut)}
\left\lbrack
\frac
{
   \theta_2(\omega t,\taut) \theta_4(\alpha x,\taux)
+i \theta_1(\omega t,\taut) \theta_3(\alpha x,\taux)
}
{
   \theta_3(\omega t,\taut) \theta_4(\alpha x,\taux)
+  \theta_4(\omega t,\taut) \theta_3(\alpha x,\taux)
}
\right\rbrack e^{-i \Omega t},
}
\\
\displaystyle{
\Omega=  \alpha^2
\left\lbrack\theta_3^4(0,\taux) + \theta_4^4(0,\taux)\right\rbrack,\
\omega=2 \alpha^2
 \frac{\theta_3^2(0,\taux)  \theta_4^2(0,\taux)}{\theta_4^2(0,\taut)},\
 \frac{\theta_3^2(0,\taut)}{\theta_4^2(0,\taut)}
=\frac{\theta_3^4(0,\taux) +\theta_4^4(0,\taux)}
    {2 \theta_3^2(0,\taux)  \theta_4^2(0,\taux)}.
}
\end{array}
\right.
\label{eqbielliptic_continuous_Theta_dark}
\end{eqnarray}
The representation in theta functions is more symmetric
and the transformation in wave numbers follows from the classical theories of
such functions \cite{AbramowitzStegun,Lawden,Chow2002}.
The expressions (\ref{eqbielliptic_continuous_sncndn_dark})
and (\ref{eqbielliptic_continuous_Theta_dark})
solve the continuous NLS in the dark soliton regime,
\begin{eqnarray}
& &
i A_t + A_{xx} -2 A^2 A^*=0.
\label{eqNLSdark}
\end{eqnarray}
In the bright soliton regime,
\begin{eqnarray}
& &
i A_t + A_{xx} +2 A^2 A^*=0,
\label{eqNLSbright}
\end{eqnarray}
the corresponding solution is
\cite{MLB1993,AA1993,Chow2002}
\begin{eqnarray}
& & \left\lbrace
\begin{array}{ll}
\displaystyle{
A=\frac{r}{\sqrt{2}}
\left\lbrack
\frac
{\left(1+\kt\right)^{-1/2}   \dn(\at t,\kt) \cn(\ax x,\kx) +i \kt^{1/2} \sn(\at t,\kt)}
{1 - \kt^{1/2}(1+k_1)^{-1/2} \cn(\at t,\kt) \cn(\ax x,\kx)}
\right\rbrack
e^{- i \Omega t},\
}
\\
\displaystyle{
\Omega=-r^2 \kt,\
\kx^2=\frac{1-\kt}{2},\
\at=\ax^2.
}
\end{array}
\right.
\label{eqbielliptic_continuous_sncndn_bright}
\end{eqnarray}
The method to obtain the above solutions
(\ref{eqbielliptic_continuous_sncndn_dark}),
(\ref{eqbielliptic_continuous_sncndn_bright})
is
either a special 
assumption \cite{MLB1993,AA1993}
or the Hirota bilinear method \cite{Chow2002}.
In the long wave limit,
the dark solution (\ref{eqbielliptic_continuous_sncndn_dark})
reduces to the collision of two dark solitons,
while the bright one (\ref{eqbielliptic_continuous_sncndn_bright})
reduces to a breather.
We shall show that both
(\ref{eqbielliptic_continuous_sncndn_dark}) and
(\ref{eqbielliptic_continuous_sncndn_bright})
have counterparts in discrete versions of NLS,
or more precisely in the AL model (\ref{eqNLSDiscreteAL}).
However, the change from continuous to discrete evolution equations is not
completely straightforward.
Even though the angular frequencies (in $t$) are still related to the wave
number (in $x$),
the constraint on the moduli of the elliptic functions will involve
the wave number as well,
see Sections \ref{sectionDiscrete_NLS-} and \ref{sectionDiscrete_NLS+}.

The paper is organized as follows.
In Section \ref{sectionDiscrete_NLS-},
we show that the defocusing discrete NLS admits
a bi-elliptic solution whose continuum limit is
(\ref{eqbielliptic_continuous_sncndn_dark}).
In Section \ref{sectionDiscrete_NLS+},
the similar solution is presented in the focusing case.
Finally,
in an Appendix,
we give in a quite symmetric notation
the continuous and discrete analytic expression
which unifies the focusing and defocusing bi-elliptic solutions
in the complex plane of $x$ and $t$.

\section     {Discrete defocusing NLS}
\label{sectionDiscrete_NLS-}

A solution for (\ref{eqNLSDiscreteAL}) in the defocusing regime
($\Sign=-1$) is
\begin{eqnarray}
& & \left\lbrace
\begin{array}{ll}
\displaystyle{
A_n=
a_0 \sqrt{\kt}
\left\lbrack
\frac{(1-\kx^2)^{1/4} \cn(\at t,\kt) +i (1-\kt^2)^{1/4} \dn(\ax n h,\kx) \sn(\at t,\kt) }
     {(1-\kx^2)^{1/4} \dn(\at t,\kt) +  (1-\kt^2)^{1/4} \dn(\ax n h,\kx)}
\right\rbrack
e^{-i \Omega t},
}
\\
\displaystyle{
a_0^2=
 \frac{\kt (1-\kx^2)^{1/2}}{h^2 (1-\kt^2)^{1/2}}
 \frac{\sn^2(\ax h,\kx)}{\dn(\ax h,k)},\
\Omega= 2 \frac{a_0^2}{\kt},\
s=      2 \frac{a_0^2}{\kt},\
}
\\
\displaystyle{
\left(\frac{1-\kx^2}{1-\kt^2}\right)^{1/2}
=1 - \frac{\kx^2}{1+\dn(\ax h ,\kx)},
}
\end{array}
\right.
\label{eqbielliptic_discreteAL_sncndn_dark}
\end{eqnarray}
in which two parameters are arbitrary, e.g.~$\ax,\kx$.
The field $A_n$ is doubly periodic in both $x=n h$ and $t$,
see Fig.~\ref{Fig1}.
A remark on the derivation is in order.
It is not clear whether a first order transformation
described earlier in the literature \cite{MLB1993,AA1993} will succeed
in the discrete version (\ref{eqNLSDiscreteAL}).
The Hirota operator relevant to discrete evolution equations will typically
involve hyperbolic functions.
Whether such an operator can be profitably applied here will be left for
future studies.
This solution (\ref{eqbielliptic_discreteAL_sncndn_dark})
is derived here by direct application of
identities between theta functions \cite{AbramowitzStegun,Lawden}.
The fact that the mere replacement of $x$ by $n h$ in
(\ref{eqbielliptic_continuous_sncndn_dark})
still yields a solution of the discrete NLS (\ref{eqNLSDiscreteAL})
just reflects the remarkable integrability properties of
(\ref{eqNLSDiscreteAL}).

\textit{The continuum limit}.
Using the expansions at the origin
\begin{eqnarray}
& &
\sn(z,k) = z + O(z^3),\
\cn(z,k) = 1 - \frac{z^2}{2} + O(z^4),\
\dn(z,k) = 1 - \frac{k^2 z^2}{2} + O(z^4),
\label{eq2.5}
\end{eqnarray}
letting $n h=x$ and taking the limit $h \to 0$,
the solution (\ref{eqbielliptic_discreteAL_sncndn_dark})
of the discrete NLS
goes to the solution
(\ref{eqbielliptic_continuous_sncndn_dark}) of the continuous NLS.

The constraint on $\kx,\kt,\ax h$ in
(\ref{eqbielliptic_discreteAL_sncndn_dark}),
which in the limit $h \to 0$ reduces
to the constraint on $\kx,\kt$ in (\ref{eqbielliptic_continuous_sncndn_dark}),
defines $\ax h$ as an elliptic integral of $(\kx,\kt)$,
and it admits real solutions if and only if $\kt < \kx$.

\textit{The long wave limit}.
In the limit $\kt \to 1$,
the solution (\ref{eqbielliptic_continuous_sncndn_dark})
of the continuous defocusing NLS (\ref{eqNLSdark})
reduces to the collision of two dark solitons.
The quantity
\begin{eqnarray}
& &
\delta=\frac{(1-\kx^2)^{1/2}}{(1-\kt^2)^{1/2}},
\end{eqnarray}
has for limit
\begin{eqnarray}
& &
\lim_{\kx \to 1,\kt \to 1} \delta=\frac{\sech \ax h}{1+\sech \ax h}.
\end{eqnarray}
Therefore the long wave limit of (\ref{eqbielliptic_discreteAL_sncndn_dark})
is the collision of two dark solitons,
\begin{eqnarray}
& & \left\lbrace
\begin{array}{ll}
\displaystyle{
A_n=A_D \frac
  {\delta \sech \at t + i \tanh \at t \sech \ax n h}
  {\delta \sech \at t +               \sech \ax n h}
e^{\displaystyle{-i \Omega_D t}},
}
\\
\displaystyle{
A_D^2=\frac{1-\sech \ax h}{h^2},\
\Omega_D= 2 A_D^2,\
\at=2 A_D^2,
}
\end{array}
\right.
\label{eqbitrigo_discreteAL_dark}
\end{eqnarray}
in which $\ax$ is arbitrary.

\section     {Discrete focusing NLS}
\label{sectionDiscrete_NLS+}

For (\ref{eqNLSDiscreteAL}) with $\Sign=1$,
starting from the known solution for the continuous case
\begin{eqnarray}
& &
A=a_0
\left\lbrack
\frac
{
   \theta_3(\omega t,\taut) \theta_2(\alpha x,\taux)
+i \theta_1(\omega t,\taut) \theta_4(\alpha x,\taux)
}
{
   \theta_4(\omega t,\taut) \theta_4(\alpha x,\taux)
-  \theta_2(\omega t,\taut) \theta_2(\alpha x,\taux)
}
\right\rbrack e^{-i \Omega t},
\label{eqbielliptic_continuous_Theta_bright}
\label{eq3.1}
\end{eqnarray}
we make in the discrete case an Ansatz consistent with that solution.
To satisfy (\ref{eqNLSDiscreteAL}),
repeated applications of theta and elliptic
functions identities now yield the solution,
\begin{eqnarray}
& & \left\lbrace
\begin{array}{ll}
\displaystyle{
A_n=
a_0
\left\lbrack
\frac
{
               \kx^{1/2}                  \dn(\at t,\kt)\cn(r n h,\kx)
+i \kt^{1/2}(1-\kx^2)^{1/4}(1-\kt^2)^{1/4}\sn(\at t,\kt)
}
{
-\kx^{1/2}\kt^{1/2} \cn(\at t,\kt)\cn(\ax n h,\kx)+ (1-\kx^2)^{1/4}(1-\kt^2)^{1/4}
}
\right\rbrack
e^{-i \Omega t}.
}
\\
\displaystyle{
a_0^2=
\frac{\kx (1-\kx^2)^{1/2}}{h^2 (1-\kt^2)^{1/2}}
 \frac{\sn^2(\ax h,\kx)}{\cn(\ax h,\kx)},\
\Omega=-2 \kt a_0^2,\
\at=2 a_0^2,
}
\\
\displaystyle{
\kx^2 + \kx \kt \left(\frac{1-\kx^2}{1-\kt^2}\right)^{1/2}
=
\frac{1}{1+ \cn(\ax h,\kx)},
}
\end{array}
\right.
\label{eqbielliptic_discreteAL_sncndn_bright}
\end{eqnarray}
in which $\ax$ and $\kx$ are arbitrary.
We have reverted back to notations of the Jacobi elliptic functions
in (\ref{eqbielliptic_discreteAL_sncndn_bright}) as they are more compact.

\textit{The continuum limit}.
Letting $n h=x$ and taking the limit $h \to 0$,
we recover the solution (\ref{eqbielliptic_continuous_sncndn_bright})
through the following steps.
We first expand  for small $h$ the dispersion relation in
(\ref{eqbielliptic_discreteAL_sncndn_bright}), which yields
\begin{eqnarray}
& &
\lim_{h \to 0} \kx^2=\frac{1-\kt}{2},
\end{eqnarray}
then, using this value of $\kx^2$,
the small $h$ expansion of the other relations in
(\ref{eqbielliptic_discreteAL_sncndn_bright})
leads to
\begin{eqnarray}
& &
\lim_{h \to 0} a_0^2= \frac{\ax^2}{2},\
\lim_{h \to 0} \Omega=- \ax^2 \kt,\
\lim_{h \to 0} \at=\ax^2.
\end{eqnarray}
Eq.~(\ref{eqbielliptic_discreteAL_sncndn_bright})
then reduces to (\ref{eqbielliptic_continuous_sncndn_bright}).

\textit{The long wave limit}.
It is of interest to take the long wave limit $\kt \to 1, \kx \to 0$
since it should yield the discrete breather.
We make use of the following limits of Jacobi elliptic functions
as $\kt \to 1, \kx \to 0$,
\begin{eqnarray}
& & \left\lbrace
\begin{array}{ll}
\displaystyle{
\dn(z, \kt) \sim \sech z,\ \dn(z, \kx) \sim 1,
}
\\
\displaystyle{
\cn(z, \kt) \sim \sech z,\ \cn(z, \kx) \sim \cos z,
}
\\
\displaystyle{
\sn(z, \kt) \sim \tanh z,\ \sn(z, \kx) \sim \sin z.
}
\end{array}
\right.
\end{eqnarray}
The two limits $\kt \to 1$ and $\kx \to 0$ are not independent
since $h$ must be kept nonzero to get the discrete breather,
and the dispersion relation in
(\ref{eqbielliptic_discreteAL_sncndn_bright}) requires
\begin{eqnarray}
& &
\frac{1}{1 + \cos \ax h}
= \frac{\kx \kt (1-\kx^2)^{1/2}}{(1-\kt)^{1/2}}
\sim \frac{\kx}{(1-\kt)^{1/2}}.
\label{eqlongwavelimit1}
\end{eqnarray}
On applying (\ref{eqlongwavelimit1})
to (\ref{eqbielliptic_discreteAL_sncndn_bright}),
we obtain
\begin{eqnarray}
& &
\lim \Omega=- \lim s=-2 \frac{1-\cos \ax h}{h^2 \cos \ax h},\
\lim a_0^2            = \frac{1-\cos \ax h}{h^2 \cos \ax h}.
\end{eqnarray}
The reality of $a_0$ implies
\begin{eqnarray}
& &
\cos \ax h> 0
\hbox{ and }
\cos \frac{\ax h}{2} \ge \frac{1}{\sqrt{2}}
\hbox{ and }
\sqrt{2} \cos \frac{\ax h}{2} - \sech \at t \cos \ax n h\not=0.
\end{eqnarray}
Applying these limits to (\ref{eqbielliptic_discreteAL_sncndn_bright}),
we have
\begin{eqnarray}
& & \left\lbrace
\begin{array}{ll}
\displaystyle{
A_n=a_0
\left\lbrack
 \frac
  { \sech \at t \cos \ax n h + i \sqrt{2} \cos(\ax h/2) \tanh \at t}
  {-\sech \at t \cos \ax n h +   \sqrt{2} \cos(\ax h/2)}
\right\rbrack
e^{-i \Omega t},
}
\\
\displaystyle{
a_0^2=\frac{1-\cos \ax h}{h^2 \cos \ax h},\
\Omega= - 2 a_0^2,\
\at= 2 a_0^2.
}
\end{array}
\right.
\label{eqbitrigo_discreteAL_sncndn_bright}
\end{eqnarray}
an expression depending on the arbitrary parameter $\ax$,
which does represent the discrete breather as expected.
This discrete breather is localized in time, see Fig.~\ref{Fig2},
as opposed to another discrete breather localized in space
\cite{CRG1995,CRG1996}.

\section{Conclusion}

A class of solutions which are doubly periodic separately in space and time
has been obtained explicitly for the integrable discrete
nonlinear Schr\"odinger equation (NLS), namely the Ablowitz-Ladik model.
Both the focusing and defocusing regimes have been treated.
These solutions, expressed as rational functions of the Jacobi elliptic
functions,
can be unified for both regimes into a single complex expression,
using a symmetric notation introduced by Halphen (Appendices A and B).

This single complex expression (\ref{eqbielliptic_discreteAL_Halphen})
is a particular instance 
of a \textit{quasiperiodic solution} (QPS)
involving four periods in the complex plane.
The mathematical problem of finding all the QPS of
the Ablowitz-Ladik equation has been solved
by various authors \cite{BP1982AL,AC1987a,MEKL},
and even beautifully extended to the whole Ablowitz-Ladik hierarchy
by Vekslerchik \cite{Vek1999}.
However, the resulting formulae are uneasy to handle for a practical use.
Our solutions,
which are derived by quite elementary algebra,
are evidently included in this abstract mathematical framework,
but they are simple \textbf{closed form} expressions,
which involve two distinct elliptic functions (genus one)
in the respective variables $n h$ and $t$.
They are the natural discretization of the solution
of Akhmediev and Ankiewicz \cite{AA1993}.

Current research efforts are made to possibly find similar solutions
for nonintegrable discretizations of the NLS
frequently encountered in physics,
as well as integrable versions of the higher order NLS.

An important continuous,
integrable higher order NLS incorporating third order dispersion
and cubic nonlinearity is the Hirota equation \cite{H1973a}.
A finite difference version has been proposed and N-soliton solutions
have been given \cite{N1990}.
An important link between the theoretical works on discrete evolution equations
and nonlinear lattices exists, as intrinsic localized modes
in a nonlinear lattice
with a hard quartic nonlinearity are governed by this discrete Hirota equation
\cite{KT1996}.
An exciting new direction for future work is thus to examine
if the doubly periodic or breather-types solutions found here can be
extended to discrete higher order NLS.
Extension and connections with nonlinear lattice dynamics can then be made.

\section*{Acknowledgments}

Partial financial support has been provided by the
Research Grants Council contracts HKU 7123/05E and
HKU 7184/04E.

\vfill\eject
\section*{Appendix A. Results in complex, symmetric notation}

\def \ax {a_x} \def \at {a_t} \def \kx {k_x} \def \kt {k_t}

In the complex plane of the variables $x$ and $t$,
the dark and bright soliton solutions,
Eqs.~(\ref{eqbielliptic_continuous_sncndn_dark})
and  (\ref{eqbielliptic_continuous_sncndn_bright}),
are not distinct
and their common expression
can be made invariant under any permutation
of three carefully defined entire or elliptic functions,
just like the Weierstrass function $\wp$
is invariant under any permutation of the three zeros $(e_1,e_2,e_3)$
of $\wp'$.
All the solutions previously given in the text
are here reexpressed in such a form.
The relevant mathematical formulae,
introduced by Halphen \cite{HalphenTraite},
can be found in Appendix B.

For convenience, let us first denote the continuous NLS,
whether in the dark regime (\ref{eqNLSdark})
or in the bright regime (\ref{eqNLSbright}),
as
\begin{eqnarray}
& &
i A_t + p A_{xx} + q A^2 A^*=0,\ p,q \in \mathcal{R},
\label{eqNLSpq}
\end{eqnarray}
and the parameters of the two elliptic functions as
\begin{eqnarray}
& & \left\lbrace
\begin{array}{ll}
\displaystyle{
\wp'(t)^2=4 (\wp(t)-E_1) (\wp(t)-E_2) (\wp(t)-E_3)
         =4 \wp^3(t)-G_2 \wp(t)-G_3,
}
\\
\displaystyle{
\wp'(x)^2=4 (\wp(x)-e_1) (\wp(x)-e_2) (\wp(x)-e_3)
         =4 \wp^3(x)-g_2 \wp(x)-g_3.
}
\end{array}
\right.
\end{eqnarray}
Finally,
let $(\alpha,\beta,\gamma)$ and $(a,b,c)$ be two independent permutations of
$(1,2,3)$.

The bi-elliptic dark or bright solution,
Eqs.~(\ref{eqbielliptic_continuous_sncndn_dark})
or (\ref{eqbielliptic_continuous_sncndn_bright}),
is then represented by the symmetric expression
\begin{eqnarray}
& & \left\lbrace
\begin{array}{ll}
\displaystyle{
A
=\frac{a_1\ha_\alpha(t)+ i a_2 \ha_b(x) \ha_\gamma(t)}
      {   \ha_\beta (t)+   a_4 \ha_b(x)}
 e^{-i \Omega t}
=\frac{a_1\sigma(x)\sigma_\alpha(t)+ i a_2 \sigma_b(x) \sigma_\gamma(t)}
      {   \sigma(x)\sigma_\beta (t)+   a_4 \sigma_b(x) \sigma       (t)}
 e^{-i \Omega t},
}
\\
\displaystyle{
a_1^2=-\frac{\Omega}{q},\
a_2^2=-\frac{2 p}{q},\
a_4  =-q a_1 a_2,\
a_4^2=2 p \Omega,\
}
\\
\displaystyle{
E_\beta -E_\gamma=\Omega^2,\
3 e_b=\frac{\Omega}{p},\
\frac{E_\alpha-E_\gamma}{E_\beta -E_\gamma}
=\left(\frac{e_a-e_c}{3 e_b}\right)^2,\
}
\end{array}
\right.
\label{eqbielliptic_continuous_Halphen}
\end{eqnarray}
in which the two arbitrary constants are, for instance,
$(e_c,\Omega)$.
The expression
(\ref{eqbielliptic_continuous_Halphen}) is invariant under
any permutation of the indices $(\alpha,\beta,\gamma)$
and $(a,b,c)$ \textit{independently}.

In the discrete case,
the solutions
(\ref{eqbielliptic_discreteAL_sncndn_dark})
and
(\ref{eqbielliptic_discreteAL_sncndn_bright})
can both be represented as,
\begin{eqnarray}
& & \left\lbrace
\begin{array}{ll}
\displaystyle{
A_n
=\frac{a_1\ha_\alpha(t)+ i a_2 \ha_b(n h) \ha_\gamma(t)}
      {   \ha_\beta (t)+   a_4 \ha_b(n h)}
 e^{-i \Omega t}
=\frac{a_1\sigma(n h)\sigma_\alpha(t)+ i a_2 \sigma_b(n h) \sigma_\gamma(t)}
      {   \sigma(n h)\sigma_\beta (t)+   a_4 \sigma_b(n h) \sigma       (t)}
 e^{-i \Omega t},
}
\\
\displaystyle{
a_1^2      =-\frac{    \Omega}{q},\
a_2^2      =-\frac{2 p}       {q h^2 \ha_a(h)\ha_c(h)},\
a_4=-q a_1 a_2,\
a_4^2      = \frac{2 p \Omega}{  h^2 \ha_a(h)\ha_c(h)},\
}
\\
\displaystyle{
E_\beta -E_\gamma=\Omega^2,\
2 \frac{\ha_a(h)\ha_c(h)-\ha_b^2(h)}{h^2 \ha_a(h)\ha_c(h)}=\frac{\Omega}{p},\
}
\\
\displaystyle{
\frac{E_\alpha-E_\gamma}{E_\beta -E_\gamma}
 =1-\frac{9 e_b^2-(e_a-e_c)^2}{4 \left(\ha_a(h)\ha_c(h)-\ha_b^2(h)\right)^2},\
}
\end{array}
\right.
\label{eqbielliptic_discreteAL_Halphen}
\end{eqnarray}
in which the two arbitrary constants are, for instance,
$(e_c,\Omega)$.
In the continuum limit $h \to 0$,
according to (\ref{eqHalphenExpansions}),
one has
\begin{eqnarray}
& &
\lim_{h \to 0} h^2 \ha_a(h)\ha_c(h)=1,\
\ha_a(h)\ha_c(h)-\ha_b^2(h)
=\frac{3 e_b}{2} - (e_a-e_c)^2 \frac{h^2}{8} + O(h^4),\
\label{eqbielliptic_limit}
\end{eqnarray}
so (\ref{eqbielliptic_discreteAL_Halphen}) goes straightforwardly to
(\ref{eqbielliptic_continuous_Halphen}).


One might wonder whether the trigonometric degeneracies
(vanishing of the elliptic discriminant)
occur
independently in $x$ and $t$ or simultaneously.
In the discrete case, the discriminants are
\begin{eqnarray}
& & {\hskip -9.0 truemm}
\left\lbrace
\begin{array}{ll}
\displaystyle{
g_2^3-27 g_3^2=16 \left[(e_b-e_c)(e_c-e_a)(e_a-e_b)\right]^2,
}
\\
\displaystyle{
G_2^3-27 G_3^2
=
256 (e_a-e_b)^2(e_b-e_c)^2
\left[(e_a-e_c)^2 - 9 e_b^2
 + 4 \left(\ha_a(h)\ha_c(h)-\ha_b^2(h)\right)^2\right]^2
\frac{a_2^4 a_4^{12}}{a_1^4}.
}
\end{array}
\right.
\nonumber
\end{eqnarray}
When the last factor in $G_2^3-27 G_3^2$ vanishes, i.e.
\begin{eqnarray}
& &
(e_a-e_c)^2 - 9 e_b^2 + 4 \left(\ha_a(h)\ha_c(h)-\ha_b^2(h)\right)^2=0,
\end{eqnarray}
by elimination of $\ha_a(h)\ha_c(h)$ and $\ha_b'(h)$ with
the rules of derivation (\ref{eqHalphenDerivation}) and
(\ref{eqHalphenDifferentialEquation}),
this implies $(e_a-e_c)\ha_b(h)=0$.
Therefore,
the trigonometric degeneracy occurs simultaneously for $x$ and $t$.
By the continuum limit,
this is also true in the continuous case,
where the two discriminants evaluate to
\begin{eqnarray}
& & \left\lbrace
\begin{array}{ll}
\displaystyle{
g_2^3-27 g_3^2=64
\left(e_3+\frac{a_1^2}{3 a_2^2} \right)^2
\left(e_3-\frac{a_1^2}{6 a_2^2} \right)^2
\left(e_3+\frac{a_1^2}{12 a_2^2} \right)^2,
}
\\
\displaystyle{
G_2^3-27 G_3^2
=
4 \left(\frac{a_2 a_4^3}{a_1 a_3^3}\right)^4
\left(e_3+\frac{a_1^2}{12 a_2^2} \right)^2 \left(g_2^3-27 g_3^2\right).
}
\end{array}
\right.
\end{eqnarray}

\vfill \eject

\section*{Appendix B. The symmetric notation of Halphen}

The notation of Jacobi
(four entire functions $\vartheta_{1,2,3,4}$
and twelve elliptic functions $\pq$, with p and q chosen among s,c,d,n)
makes the practical computations quite technical,
because of
\begin{enumerate}
\item
the lack of symmetry under a permutation of the indices
$1,2,3$ of $\vartheta_j$
or of the letters c,d,n of the twelve functions $\pq$,

\item
the lack of homogeneity,
which implies that the arguments $u,v,w$ involved in the relations between
$\wp(u), \vartheta_j(v), \pq(w)$ are all different,

\item
the high number (twelve) of elliptic functions thus defined.

\end{enumerate}

All these inconveniences have been removed by Halphen \cite{HalphenTraite},
but his elegant notation is totally absent from
the handbook \cite{AbramowitzStegun}
(the notation of Neville is different),
only presented in a less symmetric way in \cite{MOS},
and scattered among different locations in \cite{Lawden},
therefore we summarize it in the present Appendix.

In addition to the odd entire function $\sigma$ of Weierstrass,
Halphen defines three even entire functions $\sigma_\alpha,\alpha=1,2,3$,
\cite[Chap.~VIII p.~253]{HalphenTraite}
\cite[\S 10.5 p.~391]{MOS}
\cite[Eq.~(6.2.18)]{Lawden}
\begin{eqnarray}
& &
\sigma_\alpha(u)=\frac{\sigma(\omega_\alpha+u)}{\sigma(\omega_\alpha)}
                 e^{\displaystyle{-\eta_\alpha u}},\
\eta_\alpha=\zeta(\omega_\alpha),\
\alpha=1,2,3
\label{eqHalphen_sigma_alpha}
\end{eqnarray}
with the usual notation
\cite[\S 18.3.4]{AbramowitzStegun}
\begin{eqnarray}
& &
\omega_1=\omega,\ \omega_2=\omega+\omega',\ \omega_3=\omega',\
\eta_1=\eta,\ \eta_2=\eta+\eta',\ \eta_3=\eta',
\end{eqnarray}
in which $2 \omega,2 \omega'$ are the two periods.
Their symmetry with respect to permutations of the indices results
from the relation
\cite[Chap VI p.~191]{HalphenTraite}
\cite[\S 10.5 p.~391]{MOS}
\cite[Eq.~(6.9.1)]{Lawden}
\begin{eqnarray}
& &
\sqrt{\wp(u)-e_\alpha}=\frac{\sigma_\alpha(u)}{\sigma(u)},\
\alpha=1,2,3,
\label{eqHalphen_sqrtwp}
\end{eqnarray}
To avoid any ambiguity with the square roots,
let us define the three odd functions
\begin{eqnarray}
& &
\ha_\alpha(u)=\frac{\sigma_\alpha(u)}{\sigma(u)}.
\label{eqHalphen_ha_}
\end{eqnarray}
By construction, all formulae involving $\ha_\alpha$ are invariant
under any permutation $(\alpha,\beta,\gamma)$ of $(1,2,3)$.
Then,
the only relations needed to establish all the results of Appendix A are:
\hfill\break\noindent
the derivation formula
\begin{eqnarray}
& &
\ha_\alpha'(u)=-\ha_\beta(u) \ha_\gamma(u),
\label{eqHalphenDerivation}
\end{eqnarray}
the algebraic dependence relations
\begin{eqnarray}
& &
\ha_\beta^2(u)+e_\beta=\ha_\gamma^2(u)+e_\gamma,\
\label{eqHalphenDependence}
\end{eqnarray}
the addition formula
\begin{eqnarray}
& &
\ha_\alpha(u+v)=
\frac{\ha_\alpha^2(u) \ha_\alpha^2(v)-(e_\alpha-e_\beta)(e_\alpha-e_\gamma)}
{\ha_\alpha(u)\ha_\beta(v)\ha_\gamma(v)+\ha_\alpha(v)\ha_\beta(u)\ha_\gamma(u)
},
\label{eqHalphenAddition}
\end{eqnarray}
the Laurent expansion at the origin
\cite[Chap VII p.~237]{HalphenTraite} 
\cite[Chap IX  p.~304]{HalphenTraite} 
\begin{eqnarray}
& &
\ha_\alpha(u)=
\frac{1}{u}
-e_\alpha \frac{u}{2}
+\left(g_2-5 e_\alpha^2\right) \frac{u^3}{40}
+\frac{5 g_3 -7 e_\alpha g_2}{2^6.5.7} u^5
\nonumber \\ & & \phantom{xxxxx}
+\frac{28 g_2^2 - 225 e_\alpha g_3 - 105 e_\alpha^2 g_2}{2^9.3.5^2.7} u^7
+ O(u^9),
\label{eqHalphenExpansions}
\end{eqnarray}
and the degeneracy to trigonometric functions
\cite[Chap VIII (75) p.~289]{HalphenTraite}
\begin{eqnarray}
{\hskip -10.0 truemm}
& & \left\lbrace
\begin{array}{ll}
\displaystyle{
\eta \omega=\frac{\pi^2}{12},\
g_2=\frac{4}{ 3}\left(\frac{\pi}{2 \omega} \right)^4,\
g_3=\frac{8}{27}\left(\frac{\pi}{2 \omega} \right)^6,\
-\frac{e_\alpha}{2}=e_\beta=e_\gamma=-\frac{3 g_3}{2 g_2},\
}
\\
\displaystyle{
\sigma_\beta(u)=
\sigma_\gamma(u)
 =\exp \left(\displaystyle{\frac{1}{6}\left(\frac{\pi u}{2\omega}\right)^2}
       \right),\
\sigma_\alpha(u)=\sigma_\beta(u) \cos \frac{\pi u}{2 \omega},\
\sigma(u)=\frac{2 \omega}{\pi} \sigma_\beta(u) \sin \frac{\pi u}{2 \omega}.
}
\end{array}
\right.
\end{eqnarray}

From (\ref{eqHalphenDerivation}) and (\ref{eqHalphenDependence}),
one deduces the differential equation
\begin{eqnarray}
& &
\left(\ha_\alpha'(u)\right)^2=
(\ha_\alpha^2(u) + e_\alpha - e_\beta)(\ha_\alpha^2(u) + e_\alpha-e_\gamma),
\label{eqHalphenDifferentialEquation}
\end{eqnarray}
and from (\ref{eqHalphenAddition})
the expression of the three elliptic functions
$\ha_\alpha$ as rational functions of $\wp$ and $\wp'$
(a characteristic property of the elliptic functions),
\begin{eqnarray}
& &
\ha_\alpha(2 u)
=-\frac{(\wp(u,g_2,g_3)-e_\alpha)^2-(e_\alpha-e_\beta)(e_\alpha-e_\gamma)}
       {\wp'(u,g_2,g_3)}.
\end{eqnarray}

The link with the asymmetric notation of Jacobi
involves four relations between the two sets of
four entire functions, 
\cite[Chap VIII, (49) p.~260]{HalphenTraite} 
\cite[\S 10.5]{MOS}
\begin{eqnarray}
& & \left\lbrace
\begin{array}{ll}
\displaystyle{
\vartheta_1(v)=\sqrt{\frac{\omega}{\pi}} \Delta^{1/8}
{\hskip 14.8truemm}
  e^{\displaystyle{-\eta u^2/(2 \omega)}} \sigma{\hskip 0.5truemm}(u),\
u=2 \omega v,\
}
\\
\displaystyle{
\vartheta_{2}(v)=\sqrt{\frac{2 \omega}{\pi}} (e_2-e_3)^{1/4}
  e^{\displaystyle{-\eta u^2/(2 \omega)}} \sigma_1(u),\
}
\\
\displaystyle{
\vartheta_{3}(v)=\sqrt{\frac{2 \omega}{\pi}} (e_1-e_3)^{1/4}
  e^{\displaystyle{-\eta u^2/(2 \omega)}} \sigma_2(u),\
}
\\
\displaystyle{
\vartheta_{4}(v)=\sqrt{\frac{2 \omega}{\pi}} (e_1-e_2)^{1/4}
  e^{\displaystyle{-\eta u^2/(2 \omega)}} \sigma_3(u),\
}
\\
\displaystyle{
\Delta=16 (e_1-e_2)^2 (e_2-e_3)^2 (e_3-e_1)^2,\
k^2 = m = \frac{e_2 - e_3}{e_1 - e_3},\
}
\end{array}
\right.
\label{eq_thetaJacobi_function_of_sigmaHalphen}
\end{eqnarray}
plus four relations (among them a scaling one)
between the trio $\ha_\alpha$
and the copolar trio $\ps(u)$, p=c,d,n
\cite[Chap II, (16) p.~46]{HalphenTraite},
\begin{eqnarray}
& &
 \frac{\cs(z)}{\ha_1(u)}
=\frac{\ds(z)}{\ha_2(u)}
=\frac{\ns(z)}{\ha_3(u)}
=\frac{u}{z}                                   
= \frac{1}{\sqrt{e_1-e_3}}.
\label{eq_ps_ration_sigma_alpha_by_sigma}
\end{eqnarray}

\vfill \eject


\section{Figures captions}

\def \ax {a_x} \def \at {a_t} \def \kx {k_x} \def \kt {k_t}
\def \ax {r}   \def \at {s}   \def \kx {k}   \def \kt {k_1}

\begin{figure}[ht]
\caption{Intensity $|A_n|^2$ vs.~$x$ and
$t$ for the discrete defocusing bi-elliptic solution
(\ref{eqbielliptic_discreteAL_sncndn_dark}), with parameter values
$\kx=0.5,\ \kt=0.6,\ \ax=1,\ h=0.2.$ }               
\label{Fig1}
\end{figure}

\begin{figure}[ht]
\caption{
Intensity $|A_n|^2$ vs.~$x$ and $t$ for
the long wave limit
(\ref{eqbitrigo_discreteAL_sncndn_bright})
of the discrete focusing bi-elliptic solution,
with parameter values $\ax=1,\ h=0.2$.                       
}
\label{Fig2}
\end{figure}

\vfill \eject

\end{document}